\documentclass[conference,compsoc]{IEEEtran}
\IEEEoverridecommandlockouts
\usepackage{cite}
\usepackage{amsmath,amssymb,amsfonts}
\usepackage{algorithm}
\usepackage{algorithmicx}
\usepackage{algpseudocode}
\usepackage{graphicx}
\usepackage{textcomp}
\usepackage{xcolor}
\usepackage{booktabs}
\usepackage[hyphens]{url}
\usepackage[breaklinks, hidelinks]{hyperref}
\usepackage{tabularx}
\usepackage{makecell}
\usepackage{listings}
\def\BibTeX{{\rm B\kern-.05em{\sc i\kern-.025em b}\kern-.08em
    T\kern-.1667em\lower.7ex\hbox{E}\kern-.125emX}}
\begin{document}


\author{\IEEEauthorblockN{Arnabh Borah}
\IEEEauthorblockA{\textit{School of Electrical and Computer Engineering} \\
\textit{Georgia Institute of Technology}\\
Atlanta, USA \\
arnabh360@gmail.com}
\and
\IEEEauthorblockN{Md Tanvirul Alam}
\IEEEauthorblockA{\textit{Department of Computer Science} \\
\textit{Rochester Institute of Technology}\\
Rochester, USA \\
ma8235@rit.edu}
\and
\IEEEauthorblockN{Nidhi Rastogi}
\IEEEauthorblockA{\textit{Department of Computer Science} \\
\textit{Rochester Institute of Technology}\\
Rochester, USA \\
nxrvse@rit.edu}
}

\title{Adapting Large Language Models to Emerging Cybersecurity using Retrieval Augmented Generation
}
\maketitle

\begin{abstract}
Security applications are increasingly relying on large language models (LLMs) for cyber threat detection; however, their opaque reasoning often limits trust, particularly in decisions that require domain-specific cybersecurity knowledge. Because security threats evolve rapidly, LLMs must not only recall historical incidents but also adapt to emerging vulnerabilities and attack patterns. Retrieval-Augmented Generation (RAG) has demonstrated effectiveness in general LLM applications, but its potential for cybersecurity remains underexplored. In this work, we introduce a RAG-based framework designed to contextualize cybersecurity data and enhance LLM accuracy in knowledge retention and temporal reasoning. Using external datasets and the Llama-3-8B-Instruct model, we evaluate baseline RAG, an optimized hybrid retrieval approach, and conduct a comparative analysis across multiple performance metrics. Our findings highlight the promise of hybrid retrieval in strengthening the adaptability and reliability of LLMs for cybersecurity tasks.
\end{abstract}


\section{Introduction}

With the rapid advancement of cybersecurity technologies and the growing adoption of large language models (LLMs)~\cite{xu2024large,motlagh2024large,fieblinger2024actionable}, it has become imperative that LLMs adapt to evolving cyber threats and reason temporally when addressing security challenges. However, LLMs are prone to misinterpreting text, particularly when prompts contain noise or unconventional structures~\cite{madaan2022tango,alam2024ctibench}. Such misinterpretations can propagate incorrect knowledge, hindering the development of reliable reasoning strategies for future threats and increasing the likelihood of hallucinations. Moreover, the fast-paced nature of cyber incidents means that even minor reasoning errors can lead to serious consequences, making robustness and adaptability central requirements for trustworthy AI in this domain. Consequently, there is a pressing need for approaches that enable LLMs to operate effectively within the evolving landscape of cyber threat intelligence (CTI), addressing security problems over time, even under shifting semantic definitions and changing contextual requirements.

Previous benchmarking efforts on cybersecurity and CTI for LLMs have shown that, while models perform well on tasks aligned with their training data or knowledge cutoff, they often struggle to provide reliable responses when confronted with newly emerging or previously unseen information~\cite{alam2024ctibench,bhusal2024secure,ji2024sevenllm}. This limitation is particularly problematic in cybersecurity, where threats evolve rapidly and demand timely adaptation. Since full-scale pre-training or even fine-tuning on domain-specific datasets is often costly and resource-intensive~\cite{xia2024understanding}, there is a growing need for cost-efficient alternatives to ensure adaptability in security-critical domains. Retrieval-Augmented Generation (RAG)~\cite{lewis2020rag} offers a promising solution by enriching LLM responses with external, up-to-date context, thereby improving performance without the expense of retraining. Beyond efficiency, RAG also offers transparency in reasoning by grounding answers in retrieved sources, which is particularly important in security contexts where verifiable evidence is necessary. However, despite its potential, RAG has received limited attention in cybersecurity, and only a few benchmark studies have systematically evaluated its effectiveness in this domain~\cite{bhusal2024secure,rajapaksha2024rag}, leaving its practical impact underexplored.

In this paper, we make the following contributions:  
\begin{enumerate}  
    \item \textbf{Hybrid Sparse-Dense Retriever:} We propose a novel RAG framework that integrates sparse retrieval with dense semantic embedding-based similarity search, augmented with cybersecurity-specific extraction rules. This hybrid design improves contextual grounding for LLMs in CTI tasks.  

    \item \textbf{Empirical Evaluation on CTI Tasks:} We demonstrate the effectiveness of the proposed framework across multiple cybersecurity tasks, including Common Vulnerabilities and Exposures (CVEs)~\cite{cveproject2024} and Common Weakness Enumerations (CWEs)~\cite{cwe2024}. Our results show consistent improvements over both baseline LLMs and baseline RAG-augmented LLMs, utilizing the Llama-3-8B-Instruct model~\cite{llama3modelcard} and established benchmarks such as SECURE~\cite{bhusal2024secure}.  

    \item \textbf{In-depth Analysis and Guidelines:} We conduct a comprehensive analysis of key design parameters, such as temperature scaling, embedding model selection, and document extraction strategies, and provide actionable guidelines for optimizing retrieval-augmented cybersecurity reasoning in future work.  
\end{enumerate}

\section{Background}

\subsection{Related Work}

LLM evaluation in security has become an active area of research in recent years. Notable benchmarks such as CyberSecEval 2~\cite{bhatt2024cyberseceval2} and CyberSecEval 3~\cite{wan2024cyberseceval3} focus on evaluating the reactivity of LLMs under adversarial conditions, including prompt injection and related attack vectors. Other efforts, such as LLMSecCode~\cite{ryden2024llmseccode}, investigate metrics for assessing LLM performance in secure coding tasks, while datasets like CyberMetric~\cite{tihanyi2024cybermetric} and SecBench~\cite{jing2024secbench} provide thousands of multiple-choice and open-ended questions for testing cybersecurity knowledge. These resources primarily assess model accuracy and resilience, but they also highlight a recurring challenge: LLMs often struggle with reasoning when information is presented chronologically out of order~\cite{fatemi2024testoftime}. Given the continuous addition of thousands of new threats each year, such temporal reasoning limitations raise concerns about an LLM’s ability to handle older vulnerabilities once newer data is incorporated.

In parallel, Retrieval-Augmented Generation (RAG) has emerged as a promising approach for extending the knowledge and adaptability of LLMs. The survey by Gao et al.~\cite{gao2023survey} categorizes RAG methods into naive, advanced, and modular paradigms, highlighting state-of-the-art techniques across query expansion, re-ranking, and retrieval–generation fusion~\cite{setty2024improving,fan2024ragmeetsllms}. Complementary work has introduced metrics for retrieval quality~\cite{es2023ragas}, underscoring the importance of reliable document selection in downstream reasoning. Within the security domain, a limited number of studies have explored RAG-based approaches. For example, SECURE~\cite{bhusal2024secure} integrates RAG into cybersecurity benchmarking, while MORSE~\cite{simoni2024morse} introduces a RAG-based chatbot to bridge gaps in practitioner expertise. While these efforts demonstrate the potential of RAG for security applications, they remain preliminary in scope and do not systematically address hybrid retrieval strategies or temporal reasoning in CTI.

Ultimately, the integration of LLMs into cybersecurity workflows highlights both opportunities and risks. Applications such as malware detection and vulnerability triage stand to benefit from LLM-powered automation, but reliance on static knowledge bases leaves models vulnerable to newly emerging threats. Although fine-tuning or re-training on domain-specific data can mitigate this issue, such processes are computationally expensive and impractical for real-time adaptation~\cite{xia2024understanding}. This gap motivates the need for hybrid RAG frameworks tailored to cybersecurity, enabling cost-efficient adaptation while improving reliability in dynamic CTI environments.

\begin{figure}[!t]
  \centering
  \includegraphics[width=\columnwidth]{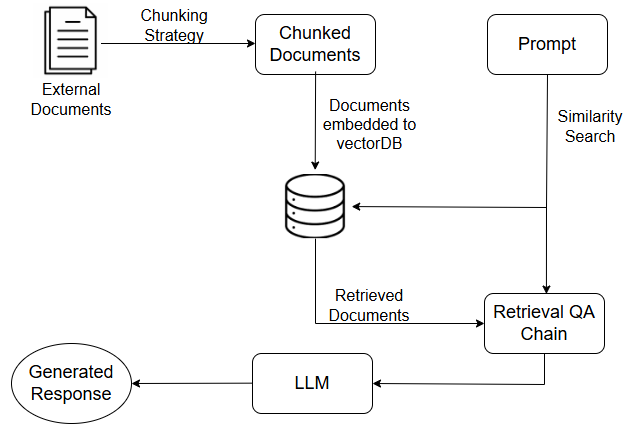}
  \caption{General overview of each major step of the RAG framework.}
  \label{fig:ragframework}
\end{figure}

\subsection{RAG Overview and Applications}

A central approach for enhancing LLM adaptability in cybersecurity is Retrieval-Augmented Generation (RAG). When training cutoffs prevent access to newly created data, retraining or fine-tuning is often costly and impractical. RAG instead provides an efficient alternative by retrieving external context to fill knowledge gaps. The framework operates by chunking documents, embedding them into a vector database (commonly with FAISS~\cite{douze2024faiss}), and retrieving the most relevant passages through similarity search. Given a user query, the top-$k$ documents are selected based on cosine similarity and combined with the query before being passed to the LLM for response generation. This mechanism allows models to interpret unfamiliar terminology and adapt to emerging threats without retraining. Figure~\ref{fig:ragframework} depicts the overall workflow, and Algorithm~\ref{alg:rag} summarizes the core steps of document chunking, embedding, retrieval, and context-augmented inference.

\begin{algorithm}[!t]
  \footnotesize
  \caption{Retrieval-Augmented Generation (RAG)}\label{alg:rag}
  \begin{algorithmic}[1]
    \Procedure{ChunkDocuments}{$filePath$}
      \State $docs \gets \Call{FileExtractor}{filePath}$
      \State $chunkedDocs \gets []$
      \ForAll{$doc$ \textbf{in} $docs$}
        \State $chunks \gets \Call{TextSplitter}{doc}$
        \State $chunkedDocs.\Call{append}{chunks}$
      \EndFor
      \State \Return $chunkedDocs$
    \EndProcedure

    \Procedure{EmbedDocuments}{$chunkedDocs$}
      \State $embedder \gets \Call{InitializeEmbedder}{modelName}$
      \State $vectorDB \gets []$
      \ForAll{$doc$ \textbf{in} $chunkedDocs$}
        \State $embedding \gets embedder.\Call{Embed}{doc}$
        \State $vectorDB.\Call{append}{(embedding, doc)}$
      \EndFor
      \State \Return $vectorDB$
    \EndProcedure

    \Procedure{RAGProcess}{$query$}
      \State $LLM \gets \Call{InitializeModel}{modelName,\,tokens,\,temp}$
      \State $chunkedDocs \gets \Call{ChunkDocuments}{path}$
      \State $vectorDB \gets \Call{EmbedDocuments}{chunkedDocs}$
      \State $retrievedDocs \gets \Call{Retrieve}{k, vectorDB, query}$
      \State $LLMAnswer \gets \Call{AnswerFromContext}{query, retrievedDocs}$
      \State \Return $LLMAnswer$
    \EndProcedure
  \end{algorithmic}
\end{algorithm}

In cybersecurity, the most relevant information sources for RAG include CVE repositories, vendor advisories, and threat intelligence platforms such as MISP~\cite{vandeplas2017misp}. Preliminary applications have demonstrated feasibility; for instance, Rajapaksha et al.~\cite{rajapaksha2024rag} built a question-answering system using the AttackER dataset~\cite{anonymous2023attacker}, while Simoni et al.~\cite{simoni2024morse} introduced a Cyber-RAG chatbot that leverages dual pipelines. These efforts demonstrate RAG’s potential to enhance LLM reasoning in CTI tasks, but they primarily focus on proof-of-concept systems rather than advancing retrieval methodologies.

At the same time, several limitations constrain RAG’s effectiveness in cybersecurity. Retrieval quality often becomes a bottleneck, with top-$k$ results excluding the correct answer even when it exists in the corpus~\cite{barnett2024failurepoints}. The accuracy and trustworthiness of retrieved context are equally critical, as noisy or malicious sources can amplify hallucinations or spread false intelligence. Moreover, cybersecurity terminology is highly nuanced (e.g., \textit{intrusion} vs.\ \textit{exploitation}, or \textit{encryption} vs.\ \textit{encoding}), making it easy for similarity-based retrievers to select superficially related but irrelevant documents. These challenges underscore the need for more robust, domain-tailored retrieval strategies that can ensure faithful context extraction and reliable downstream reasoning in dynamic CTI environments.

\section{Methodology}

We propose a hybrid retrieval framework designed to improve the accuracy and robustness of RAG in cybersecurity applications. Standard similarity-based retrieval often suffers from noisy context, ambiguous terminology, or missing critical documents when used alone. To mitigate these issues, our framework integrates dense semantic retrieval, sparse keyword-based retrieval, and cybersecurity-specific regular expression matching. This design aims to ensure that LLMs can consistently extract relevant and trustworthy context and provide accurate responses to prompts related to identifying, preventing, and mitigating threats in domains such as Common Vulnerabilities and Exposures (CVEs)~\cite{cveproject2024} and Common Weakness Enumerations (CWEs)~\cite{cwe2024}.  

\subsection{Dense Retrieval}

Dense retrieval is the standard RAG technique for capturing semantic similarity between queries and documents~\cite{karpukhin2020dense}. In our framework, we employ FAISS~\cite{douze2024faiss}, which efficiently constructs a vector database from document embeddings. At inference, the input query is encoded into an embedding vector, and a $k$-nearest neighbor (k-NN) search is performed against the FAISS index. The top-$k$ documents with the highest cosine similarity scores are returned and passed to the LLM as context. This allows the model to retrieve semantically related passages, even if the query does not exactly match the wording in the documents.

\subsection{Sparse Retrieval}

In contrast to dense methods, sparse retrieval prioritizes exact lexical matches, making it particularly useful for security identifiers such as CVE numbers. For this component, we adopt BM25~\cite{robertson2009bm25}, a widely used algorithm for keyword-based retrieval. BM25 ranks documents based on the informativeness of query terms, where frequent words (e.g., \textit{the}) are down-weighted while rare, domain-specific terms (e.g., \textit{buffer overflow}) are up-weighted. The scoring function is given by:

\begingroup
  \small
  \begin{equation}
    \mathrm{score}(D,Q)
    = \sum_{q_i\in Q}
      \mathrm{IDF}(q_i)\,
      \frac{f(q_i,D)(k_1+1)}
           {f(q_i,D)+k_1\bigl(1-b+b\,|D|/\mathrm{avgdl}\bigr)}
    \label{eq:bm25}
  \end{equation}
\endgroup

Here, $f(q_i,D)$ is the frequency of term $q_i$ in document $D$, and $\mathrm{IDF}(q_i)$ represents its informativeness. The denominator normalizes scores with respect to document length, ensuring that longer documents do not dominate the ranking~\cite{vespa2025bm25}. Unlike dense retrieval, BM25 ensures that only documents explicitly containing the query terms are returned, providing high precision for cybersecurity queries where exact identifiers are often crucial.

\begin{algorithm}[!t]
  \footnotesize
  \caption{Hybrid Sparse–Dense Retrieval Method}\label{alg:hybrid_sparse_dense}
  \begin{algorithmic}[1]
    \Procedure{RetrieveDocuments}{$query,\,k_{\mathrm{sparse}},\,k_{\mathrm{dense}},\,\alpha,$
    
    $\,\mathit{vectorDB}$}
      \State $tokenized\_q \gets query.\Call{split}{}$
      \State $scores \gets \mathit{bm25}.\Call{get\_scores}{tokenized\_q}$
      \State $sparseScores \gets \Call{sortdesc}{scores}[1{:}k_{\mathrm{sparse}}]$
      \State $top\_sparse\_indices \gets []$
      \ForAll{$score$ \textbf{in} $sparseScores$}
        \State $top\_sparse\_indices.\Call{append}{score\_index}$
      \EndFor
      \State $min\_s \gets \Call{min}{sparseScores}$
      \State $max\_s \gets \Call{max}{sparseScores}$
      \ForAll{$i$ \textbf{in} $1{:}\lvert sparseScores\rvert$}
        \State $sparseScores[i] \gets \Call{normalize}{sparseScores[i]}$
      \EndFor
      \State $sparseDocs \gets [\,chunks[i]\;\textbf{for}\;i\;\textbf{in}\;top\_sparse\_indices]$
      \State $denseDocs \gets \mathit{vectorDB}.\Call{similarity\_search}{query,$
      
      $\,k_{\mathrm{dense}}}$
      \State $denseScores \gets \mathit{vectorDB}.\Call{similarity\_scores}{query,$
      
      $\,k_{\mathrm{dense}}}$
      \State $allDocs \gets sparseDocs + denseDocs$
      \State $retrievedDocs \gets []$
      \ForAll{$doc$ \textbf{in} $allDocs$}
        \State $finalScore \gets doc.\Call{get}{sparseScore}\times\alpha$
        \State $finalScore \gets finalScore + $
        
        $(1-\alpha)\times doc.\Call{get}{denseScore}$
        \State $retrievedDocs.\Call{append}{(doc,\,finalScore)}$
      \EndFor
      \State \Return $retrievedDocs$
    \EndProcedure
  \end{algorithmic}
\end{algorithm}

\subsection{Hybrid Sparse–Dense Retriever}

Although both dense and sparse retrievals have strengths, they also have complementary weaknesses: dense retrieval may surface semantically similar but irrelevant passages, while sparse retrieval may miss semantically aligned documents that lack exact keywords. To address this, we propose a Hybrid Sparse–Dense Retriever that integrates the two.  

A weighting parameter $\alpha$ determines the contribution of each retriever to the final score:
\begin{equation}
\mathrm{score}(D_i) = \alpha \cdot \mathrm{score}_h(D_i) + (1 - \alpha) \cdot \mathrm{score}_d(D_i),
\label{finalScore}
\end{equation}
where $\mathrm{score}_h(D_i)$ and $\mathrm{score}_d(D_i)$ denote the BM25 and dense similarity scores for document $D_i$. A higher $\alpha$ favors sparse retrieval, while a lower $\alpha$ emphasizes semantic similarity.  

Since BM25 scores are unbounded, while dense scores lie in $[0,1]$, Min–Max Normalization is applied to align the scales. The hybrid retrieval algorithm (Algorithm~\ref{alg:hybrid_sparse_dense}) accepts $k_{\mathrm{sparse}}$ and $k_{\mathrm{dense}}$ as parameters, retrieves candidates from both retrievers, normalizes and weights scores, and outputs a ranked set of documents for the LLM.

\subsection{Regular Expression Matching for CVEs}

To further tailor the framework to cybersecurity tasks, we integrate a regular-expression-based filter for CVE identifiers. When a query includes a CVE-ID, this identifier typically serves as the strongest signal for relevant document retrieval. We therefore apply the regex pattern [r`CVE-[0-9]{4}-[0-9]{4,6}`] from LADDER~\cite{aiforsec_heuristics_ner} to identify matching documents. Each match receives a score boost of $+1.0$, ensuring high recall for vulnerability-specific queries.  

This mechanism significantly improves retrieval precision when users provide CVE-IDs explicitly and ensures that the framework can surface reliable context even when minimal information is available. For example, if the only input is a CVE string, the system can still retrieve accurate descriptions, advisories, and mitigations, thereby improving LLM performance in time-sensitive security scenarios.

\section{Experimental Setting}

We evaluate our proposed framework using the KCV and CWET datasets from the SECURE benchmark~\cite{bhusal2024secure}. These datasets are widely used for assessing LLM performance in cybersecurity reasoning, particularly for tasks involving vulnerability and weakness identification. For all experiments, we adopt the Llama-3-8B-Instruct model~\cite{llama3modelcard} as the base LLM. The external context for RAG is derived from official cybersecurity sources, including CVE descriptors (with a focus on 2024 entries)~\cite{cveproject2024} and the CWE knowledge base~\cite{cwe2024}.  

\subsection{Baseline Without RAG}

\begin{figure}[!t]
  \centering
  \includegraphics[width=0.82\columnwidth]{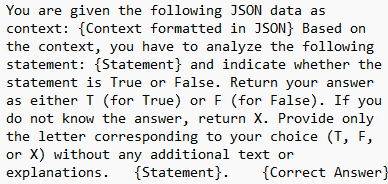}
  \caption{General prompt format in the KCV dataset for non-RAG evaluation.}
  \label{fig:promptformat}
\end{figure}

We first establish baselines by evaluating the model without any retrieval augmentation. In this setting, the temperature parameter is fixed at the default value of $0.7$, and the output is restricted to a single generated token per question to ensure deterministic comparisons. The prompts follow the exact format defined in the SECURE-KCV benchmark~\cite{bhusal2024secure}, enabling direct comparability with prior results. Figure~\ref{fig:promptformat} illustrates the prompt structure for KCV questions under non-RAG evaluation, while CWET prompts follow a similar format but consist of four multiple-choice options (A–D) provided as the only possible answers with no additional context. For evaluation, the complete KCV dataset of 466 questions was used, whereas the CWET dataset was reduced to 103 questions after filtering.

\subsection{Baseline RAG}
We next incorporate the baseline RAG framework (Algorithm~\ref{alg:rag}) to provide the LLM with external context during evaluation. For KCV, relevant context was retrieved from the CVE repository~\cite{cveproject2024}, while CWET relied on descriptions from the CWE knowledge base. The framework used the \texttt{RecursiveCharacterTextSplitter} function from LangChain~\cite{langchain2024}, with a chunk size of 512 characters and an overlap of 20 characters, to segment documents into retrievable units. For embeddings, we employed the \texttt{mixedbread-ai/mxbai-embed-large-v1} model~\cite{lee2024mxbai}, and the resulting vectors were indexed using FAISS~\cite{facebook2017faiss}. During inference, the \texttt{ConversationalRetrievalChain} retrieved the top-$k$ documents (defaulted to 3) and passed them, alongside the original query, to the LLM for answer generation.

\subsection{Hybrid Sparse–Dense Retrieval}

Finally, we evaluate our hybrid sparse–dense retriever, which builds on the baseline RAG setup by incorporating BM25 keyword retrieval alongside FAISS-based semantic retrieval and combining their results using the weighted scoring method described in Section~\ref{alg:hybrid_sparse_dense}. To ensure comparability, the same experimental parameters were retained across settings. For KCV, both sparse and dense scores were integrated. In contrast, for CWET, the regex-based CVE matcher was omitted because most questions lacked explicit CWE identifiers, and evaluation relied solely on the hybrid retriever. This design isolates the contributions of sparse retrieval and hybrid weighting while maintaining consistency with prior baselines.

\subsection{Evaluation Protocol}
Across all settings, experiments were repeated multiple times to reduce variance, and the results were averaged to report stable accuracy values. Metrics focused on the correctness of the model’s response to each prompt, comparing performance across (i) baseline LLM without RAG, (ii) baseline RAG, and (iii) the proposed hybrid framework. This experimental pipeline enables a direct assessment of whether retrieval strategies improve the reliability of LLMs in CTI reasoning tasks.

\section{Results and Discussion}
\begin{table}[t]
  \centering
  \caption{Results for each Setting}
  \begin{tabular}{lcc}
    \toprule
    Setting              & KCV accuracy (\%)         & CWET accuracy (\%) \\ & mean $\pm \hspace{0.5 mm} \sigma$ & mean $\pm \hspace{0.5 mm} \sigma$\\
    \midrule
    No RAG               & $59.2\pm0$       & $85.4\pm0$ \\
    \midrule
    Preformatted Context               & $82.8\pm0.48$       & — \\
    \midrule
    Baseline RAG         & $57.6\pm1.16$       & $86.4\pm0$ \\
    \midrule
    Full Hybrid         & $62.5\pm0.78$       & $92.2\pm0$               \\
    \midrule
    Full Hybrid + Regex  & $72.7\pm1.37$       & — \\
    \bottomrule
  \end{tabular}
  \label{tab1}
\end{table}

Table~\ref{tab1} summarizes the KCV results across all evaluation settings, reporting both mean accuracy and standard deviation ($\sigma$). All results are averaged over ten iterations. Without any contextual support, the LLM achieved 59.2\% accuracy and consistently predicted `F' (false) for every question, highlighting its limitations in handling unseen vulnerabilities. When supplied with pre-formatted and manually curated context, accuracy improved significantly to 82.8\%, demonstrating the impact of structured, high-quality evidence. In contrast, the baseline RAG configuration underperformed at 57.6\%, suggesting that naïve retrieval can introduce noisy or misleading context. The hybrid sparse–dense retriever provided a modest gain, reaching 62.5\%, while the hybrid model enhanced with regex achieved the best performance of 72.7\% in a fully automated setting, narrowing the gap toward curated context. The low standard deviations across runs indicate that these improvements are consistent and robust.  

Turning to CWET, the hybrid model once again performed best, attaining an accuracy of 92.2\%. Given that CWET questions are multiple-choice with four options, random guessing would yield only 25\% accuracy, underscoring the strength of all models in this setting. Interestingly, the LLM without RAG already achieved 85.4\%, suggesting that Llama-3-8B-Instruct may have been exposed to CWE-related knowledge during pre-training. The baseline RAG setup provided only a marginal increase to 86.4\%, whereas the hybrid approach delivered a meaningful boost of nearly 7 percentage points.  

For the KCV dataset, however, errors remained more pronounced: the hybrid model produced an average of 130 incorrect responses out of 466 questions. Moreover, specific questions were consistently misclassified across runs, suggesting systematic weaknesses rather than random variation. One representative example is provided below, along with the three retrieved documents used for context, illustrating how formatting noise and terminology mismatches can mislead the model despite improved retrieval.

\begin{lstlisting}[language=HTML] 
Prompt: `The vulnerability described in CVE-2024-5022 affects Mozilla Firefox for iOS versions less than 126.' -> A: `F'

Document 1: The file schemes of URLs would be hidden, resulting in potential spoofing of a website's address in the location bar This vulnerability affects Focus for iOS < 126.Affected: Mozilla Focus for iOSVersion unspecified: affectedProblemType: None URLswith file scheme could have been used to spoof adresses in the location barReference():

Document 2: By supplying local file paths, such as `file:///etc/passwd', an attacker can read sensitive data from the server. This vulnerability arises because the system does not properly validate or sanitize the user input from the URL field. Speicifcally: 1. The URL input (`<input data-v-5fc86d7="" id="url" type="url" class="form-control" pattern="https?://.+" required="">') allows users to input arbitrary file paths,

Document 3: Any outside party can read them with an unauthorized HTTP GET request to the same server hosting the rest of the website. While some server files are hashed, making their access obscure, the files corresponding to the file system router (those in `src/pages') are predictably named. For example, the sourcemap file for `src/pages/index.astro' gets named  `dist/client/pages/index.astro.mjs.map'. This vulnerability is the root cause of issue #12703, which links to a
\end{lstlisting}

\begin{table*}[!t]
  \centering
  \footnotesize
  \caption{KCV Dataset Accuracy for Llama-3-8B-Instruct Model under Varying Temperatures}
  \label{tab:temps}
  \setlength{\tabcolsep}{8pt} 
  \begin{tabularx}{\textwidth}{>{\raggedright\arraybackslash}X 
                              *{3}{>{\centering\arraybackslash}X}}
    \hline
    Evaluation Setting            & Accuracy ($\%$) & Accuracy ($\%$) & Accuracy ($\%$) \\ & (Temp = 0.01) & (Temp = 0.7) & (Temp = 1.0) \\
    & mean $\pm \hspace{0.5 mm} \sigma$ & mean $\pm \hspace{0.5 mm} \sigma$ & mean $\pm \hspace{0.5 mm} \sigma$\\
    \hline
    LLM with Preformatted Context     & $84.4 \pm 0.26$             & $82.8 \pm 0.48$            & $81.3 \pm 1.12$            \\
    LLM with Full Hybrid RAG                     & $67.3 \pm 0.27$             & $62.5 \pm 0.78$            & $57.1 \pm 2.02$            \\
    LLM with Full Hybrid + Regex                 & $76.3 \pm 0.36$             & $72.7 \pm 1.37$            & $67.2 \pm 2.13$            \\
    \hline
  \end{tabularx}
  \label{tab2}
\end{table*}

\begin{table*}[!t]
  \centering
  \footnotesize
  \caption{Performance Accuracy of Various Embedding Models on the CWET Dataset}
  \label{tab:embedding_performance}
  \setlength{\tabcolsep}{4pt} 
  \begin{tabularx}{\textwidth}{*{6}{>{\centering\arraybackslash}X}}
    \hline
    Model
      & \makecell[l]{mixedbread-ai/\\mxbai-embed-\\large-v1}
      & \makecell[l]{sentence-\\transformers/\\all-mpnet-base-v2}
      & \makecell[l]{sentence-\\transformers/\\multi-qa-MiniLM-\\L6-dot-v1}
      & \makecell[l]{sentence-\\transformers/\\msmarco-distilbert\\-base-v3}
      & \makecell[l]{hkunlp/\\instructor-large} \\
    \hline
    Accuracy (\%)\\
    mean $\pm \hspace{0.5 mm} \sigma$
      & \hspace{-1.1 cm}$92.23 \pm 0$ 
      & \hspace{-1.1 cm}$88.35 \pm 0$ 
      & \hspace{-1.1 cm}$90.29 \pm 0$
      & \hspace{-1.1 cm}$91.26 \pm 0$ 
      & \hspace{-1.1 cm}$89.32 \pm 0$  \\
    \hline
  \end{tabularx}
  \label{tab3}
\end{table*}

From the retrieved evidence, Document~1 explicitly states \textit{“This vulnerability affects \ldots iOS $<$ 12.6”}, which confirms the correct answer as true. Nevertheless, the LLM outputs `F' (false). This error is likely caused by noise in the retrieved context: the lack of sentence boundaries makes the supporting evidence more difficult to parse, and the presence of the word \textit{Focus} may have led the model to treat “Firefox Focus” as unrelated to Firefox.  

Such mistakes are common in KCV questions, particularly when CVE IDs are included in the prompt. These observations reinforce the hypothesis that LLMs are highly sensitive to context formatting and also explain why pre-formatted, manually curated context yields the highest accuracy, even though such settings are less generalizable compared to automated retrieval. Improving the clarity and structure of retrieved passages is therefore an essential step toward enhancing RAG performance on vulnerability-related tasks. Even minor formatting inconsistencies can cause disproportionate errors, highlighting the need for retrieval pipelines that not only select the correct documents but also present them in a model-friendly format.  

By contrast, CWET documents are provided in PDF form, with complete sentences and richer contextual descriptions. This structural difference reduces ambiguity, resulting in nearly identical performance for the LLM with and without the baseline RAG. However, the hybrid RAG model achieves a further 7\% improvement, which we attribute to BM25’s ability to exploit longer, well-formed text windows. In this case, keyword matching complements semantic retrieval more effectively, suggesting that hybrid retrieval is helpful when document quality and structure vary across CTI datasets.

\section{Ablation Studies}

\subsection{Temperature Setting}
We first examine the impact of the temperature parameter on model accuracy. Three values were tested: 0.01, 0.7 (default), and 1.0. Table~\ref{tab2} shows how performance varied under these settings. Across all configurations, accuracy was consistently higher at lower temperatures, indicating that a more deterministic decoding strategy is beneficial for cybersecurity reasoning tasks. In contrast, increasing the temperature led to increased variability and less reliable outputs, often resulting in random responses. These findings suggest that creativity-oriented sampling is detrimental in this domain, where factual precision and consistency are critical.

\subsection{Embedding Models}
We also evaluate the effect of different embedding models on retrieval quality. Five models were compared: the default `mixedbread-ai/mxbai-embed-large-v1'~\cite{lee2024mxbai}, `sentence-transformers/all-mpnet-base-v2'~\cite{sentence-transformers-all-mpnet-base-v2}, `sentence-transformers/multi-qa-MiniLM-L6-dot-v1'~\cite{st-multiqa-minilm-l6-dot-v1}, `sentence-transformers/msmarco-distilbert-base-v3'~\cite{st-msmarco-distilbert-base-v3}, and `hkunlp/instructor-large'~\cite{hkunlp-instructor-large}. Table~\ref{tab3} reports their performance on the CWET dataset. Overall, the results show relatively small differences across models: while `mxbai-embed-large-v1' consistently achieved the highest accuracy, its advantage over the much smaller `multi-qa-MiniLM-L6-dot-v1' was only 1.94\%. Given that the former has 335M parameters compared to just 22.7M for the latter, this trade-off highlights the importance of resource constraints. In practice, the choice of embedding model may depend less on absolute accuracy and more on deployment requirements, such as memory footprint, inference speed, or scalability within large-scale CTI pipelines.

\section{Limitations and Future Work}

While our hybrid model demonstrates promising improvements, several limitations remain. First, the framework has been evaluated so far only on the KCV and CWET datasets, both of which focus on CVE and CWE contexts. To improve generalization, future work should test the approach in a broader range of security-related datasets, including those without explicit identifiers and those with short-answer or free-text formats. Expanding evaluation to other CTI datasets, such as threat advisories or malware reports, could help assess the framework's robustness to different data distributions.

Second, the current regex matcher is limited to CVE patterns. Extending this capability to include other identifiers such as CWE, ATT\&CK, CAPEC, and additional vulnerability taxonomies could further enhance retrieval precision. Beyond identifier-based improvements, measuring faithfulness to retrieved context and developing more reliable retrieval metrics remain critical open challenges. Future studies could also explore automated faithfulness measures and retrieval calibration methods to better quantify the model's trustworthiness.

Finally, the current study is restricted to Llama-3-8B-Instruct. Evaluating the framework across different model families and scales would provide more substantial evidence of the generalizability of the hybrid sparse–dense retriever. Longer-term directions include exploring lightweight postprocessing of retrieved context, integrating multimodal retrieval (for example, diagrams or vendor advisories), and assessing the feasibility of deployment in real-world CTI pipelines with respect to latency, cost, and reliability.

\section{Conclusion}

This paper introduced a hybrid sparse–dense retriever to address the inconsistency of standard RAG in cybersecurity, combining BM25 keyword matching with FAISS-based semantic retrieval and augmenting it with regular expression matching for CVE identifiers. Our experiments show that this framework consistently improves context retrieval and LLM accuracy compared to baseline methods. While the current evaluation is limited to CVE- and CWE-focused datasets, future work should extend the approach to broader security contexts, additional identifiers such as CWE and ATT\&CK, and diverse model architectures. We believe these findings highlight the importance of robust retrieval strategies in advancing the integration of RAG into real-world CTI systems.

\section*{Acknowledgment}

The first author would like to thank the research experience for undergraduates (REU) program at Rochester Institute of Technology for the opportunity to conduct this research. This material is based upon work supported by the National
Science Foundation Award 2447631: Trustworthy AI. Any opinions, findings, and conclusions or recommendations expressed
in this material are those of the author(s) and do not necessarily
reflect the views of the National Science Foundation.



\clearpage

\bibliographystyle{IEEEtran}
\bibliography{refs}

\begin{thebibliography}{10}
\providecommand{\url}[1]{#1}
\csname url@samestyle\endcsname
\providecommand{\newblock}{\relax}
\providecommand{\bibinfo}[2]{#2}
\providecommand{\BIBentrySTDinterwordspacing}{\spaceskip=0pt\relax}
\providecommand{\BIBentryALTinterwordstretchfactor}{4}
\providecommand{\BIBentryALTinterwordspacing}{\spaceskip=\fontdimen2\font plus
\BIBentryALTinterwordstretchfactor\fontdimen3\font minus \fontdimen4\font\relax}
\providecommand{\BIBforeignlanguage}[2]{{%
\expandafter\ifx\csname l@#1\endcsname\relax
\typeout{** WARNING: IEEEtran.bst: No hyphenation pattern has been}%
\typeout{** loaded for the language `#1'. Using the pattern for}%
\typeout{** the default language instead.}%
\else
\language=\csname l@#1\endcsname
\fi
#2}}
\providecommand{\BIBdecl}{\relax}
\BIBdecl

\bibitem{xu2024large}
H.~Xu, S.~Wang, N.~Li, K.~Wang, Y.~Zhao, K.~Chen, T.~Yu, Y.~Liu, and H.~Wang, ``Large language models for cyber security: A systematic literature review,'' \emph{arXiv preprint arXiv:2405.04760}, 2024.

\bibitem{motlagh2024large}
F.~N. Motlagh, M.~Hajizadeh, M.~Majd, P.~Najafi, F.~Cheng, and C.~Meinel, ``Large language models in cybersecurity: State-of-the-art,'' \emph{arXiv preprint arXiv:2402.00891}, 2024.

\bibitem{fieblinger2024actionable}
R.~Fieblinger, M.~T. Alam, and N.~Rastogi, ``Actionable cyber threat intelligence using knowledge graphs and large language models,'' in \emph{2024 IEEE European symposium on security and privacy workshops (EuroS\&PW)}.\hskip 1em plus 0.5em minus 0.4em\relax IEEE, 2024, pp. 100--111.

\bibitem{madaan2022tango}
A.~Madaan and A.~Yazdanbakhsh, ``Text and patterns: For effective chain of thought, it takes two to tango,'' \emph{arXiv preprint arXiv:2209.07686}, 2022.

\bibitem{alam2024ctibench}
M.~T. Alam, D.~Bhusal, L.~Nguyen, and N.~Rastogi, ``Ctibench: A benchmark for evaluating llms in cyber threat intelligence,'' \emph{Advances in Neural Information Processing Systems}, vol.~37, pp. 50\,805--50\,825, 2024.

\bibitem{bhusal2024secure}
\BIBentryALTinterwordspacing
D.~Bhusal, M.~T. Alam, L.~Nguyen, A.~Mahara, Z.~Lightcap, R.~Frazier, R.~Fieblinger, G.~L. Torales, B.~A. Blakely, and N.~Rastogi, ``Secure: Benchmarking large language models for cybersecurity,'' \emph{arXiv preprint arXiv:2405.20441}, 2024. [Online]. Available: \url{https://arxiv.org/abs/2405.20441}
\BIBentrySTDinterwordspacing

\bibitem{ji2024sevenllm}
H.~Ji, J.~Yang, L.~Chai, C.~Wei, L.~Yang, Y.~Duan, Y.~Wang, T.~Sun, H.~Guo, T.~Li \emph{et~al.}, ``Sevenllm: Benchmarking, eliciting, and enhancing abilities of large language models in cyber threat intelligence,'' \emph{arXiv preprint arXiv:2405.03446}, 2024.

\bibitem{xia2024understanding}
Y.~Xia, J.~Kim, Y.~Chen, H.~Ye, S.~Kundu, C.~C. Hao, and N.~Talati, ``Understanding the performance and estimating the cost of llm fine-tuning,'' in \emph{2024 IEEE International Symposium on Workload Characterization (IISWC)}.\hskip 1em plus 0.5em minus 0.4em\relax IEEE, 2024, pp. 210--223.

\bibitem{lewis2020rag}
\BIBentryALTinterwordspacing
P.~Lewis, E.~Perez, A.~Piktus, F.~Petroni, V.~Karpukhin, N.~Goyal, H.~K{\"u}ttler, M.~Lewis, W.~tau Yih, T.~Rockt{\"a}schel, S.~Riedel, and D.~Kiela, ``Retrieval-augmented generation for knowledge-intensive {NLP} tasks,'' 2020. [Online]. Available: \url{https://arxiv.org/abs/2005.11401}
\BIBentrySTDinterwordspacing

\bibitem{rajapaksha2024rag}
\BIBentryALTinterwordspacing
S.~Rajapaksha, R.~Rani, and E.~Karafili, ``A rag-based question-answering solution for cyber-attack investigation and attribution,'' 2024, accepted at SECAI 2024 (ESORICS 2024). [Online]. Available: \url{https://arxiv.org/abs/2408.06272}
\BIBentrySTDinterwordspacing

\bibitem{cveproject2024}
\BIBentryALTinterwordspacing
{CVE Project}. (2024) Cves published in 2024. [Online]. Available: \url{https://github.com/CVEProject/cvelistV5/tree/main/cves/2024}
\BIBentrySTDinterwordspacing

\bibitem{cwe2024}
\BIBentryALTinterwordspacing
{MITRE CWE}. (2024) A community-developed list of sw \& hw weaknesses that can become vulnerabilities. [Online]. Available: \url{https://cwe.mitre.org/index.html}
\BIBentrySTDinterwordspacing

\bibitem{llama3modelcard}
\BIBentryALTinterwordspacing
AI@Meta, ``Llama 3 model card,'' 2024. [Online]. Available: \url{https://github.com/meta-llama/llama3/blob/main/MODEL_CARD.md}
\BIBentrySTDinterwordspacing

\bibitem{bhatt2024cyberseceval2}
\BIBentryALTinterwordspacing
M.~Bhatt, S.~Chennabasappa, Y.~Li, C.~Nikolaidis, D.~Song, S.~Wan, F.~Ahmad, C.~Aschermann, Y.~Chen, D.~Kapil, D.~Molnar, S.~Whitman, and J.~Saxe, ``Cyberseceval 2: A wide‐ranging cybersecurity evaluation suite for large language models,'' \emph{arXiv preprint arXiv:2404.13161}, 2024. [Online]. Available: \url{https://arxiv.org/abs/2404.13161}
\BIBentrySTDinterwordspacing

\bibitem{wan2024cyberseceval3}
\BIBentryALTinterwordspacing
S.~Wan, C.~Nikolaidis, D.~Song, D.~Molnar, J.~Crnkovich, J.~Grace, M.~Bhatt, S.~Chennabasappa, S.~Whitman, S.~Ding, V.~Ionescu, Y.~Li, and J.~Saxe, ``Cyberseceval 3: Advancing the evaluation of cybersecurity risks and capabilities in large language models,'' \emph{arXiv preprint arXiv:2408.01605}, 2024. [Online]. Available: \url{https://arxiv.org/abs/2408.01605}
\BIBentrySTDinterwordspacing

\bibitem{ryden2024llmseccode}
\BIBentryALTinterwordspacing
A.~Ryd{\'e}n, E.~N{\"a}slund, E.~M. Schiller, and M.~Almgren, ``Llmseccode: Evaluating large language models for secure coding,'' \emph{arXiv preprint arXiv:2408.16100}, 2024. [Online]. Available: \url{https://arxiv.org/abs/2408.16100}
\BIBentrySTDinterwordspacing

\bibitem{tihanyi2024cybermetric}
\BIBentryALTinterwordspacing
N.~Tihanyi, M.~Ferrag, R.~Jain, T.~Bisztray, and M.~Debbah, ``Cybermetric: A benchmark dataset based on retrieval‑augmented generation for evaluating llms in cybersecurity knowledge,'' \emph{arXiv preprint arXiv:2402.07688}, 2024. [Online]. Available: \url{https://arxiv.org/abs/2402.07688}
\BIBentrySTDinterwordspacing

\bibitem{jing2024secbench}
\BIBentryALTinterwordspacing
P.~Jing, M.~Tang, X.~Shi, X.~Zheng, S.~Nie, S.~Wu, Y.~Yang, and X.~Luo, ``Secbench: A comprehensive multi‑dimensional benchmarking dataset for llms in cybersecurity,'' \emph{arXiv preprint arXiv:2412.20787}, 2024. [Online]. Available: \url{https://arxiv.org/abs/2412.20787}
\BIBentrySTDinterwordspacing

\bibitem{fatemi2024testoftime}
\BIBentryALTinterwordspacing
B.~Fatemi, M.~Kazemi, A.~Tsitsulin, K.~Malkan, J.~Yim, J.~Palowitch, S.~Seo, J.~Halcrow, and B.~Perozzi, ``Test of time: A benchmark for evaluating llms on temporal reasoning,'' \emph{arXiv preprint arXiv:2406.09170}, 2024. [Online]. Available: \url{https://arxiv.org/abs/2406.09170}
\BIBentrySTDinterwordspacing

\bibitem{gao2023survey}
\BIBentryALTinterwordspacing
Y.~Gao, Y.~Xiong, X.~Gao, K.~Jia, J.~Pan, Y.~Bi, Y.~Dai, J.~Sun, M.~Wang, and H.~Wang, ``Retrieval‑augmented generation for large language models: A survey,'' 2023, arXiv:2312.10997. [Online]. Available: \url{https://arxiv.org/abs/2312.10997}
\BIBentrySTDinterwordspacing

\bibitem{setty2024improving}
\BIBentryALTinterwordspacing
S.~Setty, H.~Thakkar, A.~Lee, E.~Chung, and N.~Vidra, ``Improving retrieval for {RAG} based question answering models on financial documents,'' \emph{arXiv preprint arXiv:2404.07221}, 2024. [Online]. Available: \url{https://arxiv.org/abs/2404.07221}
\BIBentrySTDinterwordspacing

\bibitem{fan2024ragmeetsllms}
W.~Fan, Y.~Ding, L.~Ning, S.~Wang, H.~Li, D.~Yin, T.-S. Chua, and Q.~Li, ``A survey on rag meeting llms: Towards retrieval-augmented large language models,'' in \emph{Proceedings of the 30th ACM SIGKDD conference on knowledge discovery and data mining}, 2024, pp. 6491--6501.

\bibitem{es2023ragas}
\BIBentryALTinterwordspacing
S.~Es, J.~James, L.~Espinosa‑Anke, and S.~Schockaert, ``Ragas: Automated evaluation of retrieval augmented generation,'' \emph{arXiv preprint arXiv:2309.15217}, 2023. [Online]. Available: \url{https://arxiv.org/abs/2309.15217}
\BIBentrySTDinterwordspacing

\bibitem{douze2024faiss}
\BIBentryALTinterwordspacing
M.~Douze, A.~Guzhva, C.~Deng, J.~Johnson, G.~Szilvasy, P.-E. Mazar{\'e}, M.~Lomeli, L.~Hosseini, and H.~J{\'e}gou, ``The faiss library,'' \emph{arXiv preprint arXiv:2401.08281}, 2024. [Online]. Available: \url{https://arxiv.org/abs/2401.08281}
\BIBentrySTDinterwordspacing

\bibitem{vandeplas2017misp}
C.~Vandeplas, A.~Iklody \emph{et~al.}, ``Misp threat sharing: Malware information sharing platform \& threat sharing,'' \url{https://www.misp-project.org/}, 2017, accessed: 2025-07-19.

\bibitem{anonymous2023attacker}
\BIBentryALTinterwordspacing
{Anonymous}, ``{AttackER: NER Attack Attribution},'' 2023. [Online]. Available: \url{https://zenodo.org/records/10276922}
\BIBentrySTDinterwordspacing

\bibitem{barnett2024failurepoints}
\BIBentryALTinterwordspacing
S.~Barnett, S.~Kurniawan, S.~Thudumu, Z.~Brannelly, and M.~Abdelrazek, ``Seven failure points when engineering a retrieval augmented generation system,'' \emph{arXiv preprint arXiv:2401.05856}, 2024. [Online]. Available: \url{https://arxiv.org/abs/2401.05856}
\BIBentrySTDinterwordspacing

\bibitem{karpukhin2020dense}
\BIBentryALTinterwordspacing
V.~Karpukhin, B.~Oguz, S.~Min, P.~Lewis, L.~Wu, S.~Edunov, and W.-t. Yih, ``Dense passage retrieval for open-domain question answering,'' in \emph{Proc.\ of the 2020 Conference on Empirical Methods in Natural Language Processing (EMNLP)}, 2020, pp. 6769--6781. [Online]. Available: \url{https://arxiv.org/abs/2004.04906}
\BIBentrySTDinterwordspacing

\bibitem{robertson2009bm25}
S.~Robertson and H.~Zaragoza, ``The probabilistic relevance framework: Bm25 and beyond,'' \emph{Foundations and Trends® in Information Retrieval}, vol.~3, no.~4, pp. 333--389, 2009.

\bibitem{vespa2025bm25}
\BIBentryALTinterwordspacing
{Vespa.ai}. (2025) Bm25 reference. [Online]. Available: \url{https://docs.vespa.ai/en/reference/bm25.html}
\BIBentrySTDinterwordspacing

\bibitem{aiforsec_heuristics_ner}
{AIforSec}, ``{heuristics\_ner.py},'' \url{https://github.com/aiforsec/LADDER/blob/main/ner/heuristics_ner.py}, 2024, accessed: 2025-07-19.

\bibitem{langchain2024}
\BIBentryALTinterwordspacing
{LangChain}. (2024) Introduction. [Online]. Available: \url{https://python.langchain.com/v0.2/docs/introduction/}
\BIBentrySTDinterwordspacing

\bibitem{lee2024mxbai}
\BIBentryALTinterwordspacing
S.~Lee, A.~Shakir, D.~Koenig, and J.~Lipp. (2024) Open source strikes bread—new fluffy embeddings model. [Online]. Available: \url{https://www.mixedbread.ai/blog/mxbai-embed-large-v1}
\BIBentrySTDinterwordspacing

\bibitem{facebook2017faiss}
\BIBentryALTinterwordspacing
{Facebook AI}. (2017) Faiss: A library for efficient similarity search by meta. [Online]. Available: \url{https://engineering.fb.com/2017/03/29/data-infrastructure/faiss-a-library-for-efficient-similarity-search/}
\BIBentrySTDinterwordspacing

\bibitem{sentence-transformers-all-mpnet-base-v2}
\BIBentryALTinterwordspacing
{Sentence‑Transformers}. (2021) all‑mpnet‑base‑v2. [Online]. Available: \url{https://huggingface.co/sentence-transformers/all-mpnet-base-v2}
\BIBentrySTDinterwordspacing

\bibitem{st-multiqa-minilm-l6-dot-v1}
\BIBentryALTinterwordspacing
------. (2021) multi‑qa‑minilm‑l6‑dot‑v1. [Online]. Available: \url{https://huggingface.co/sentence-transformers/multi-qa-MiniLM-L6-dot-v1}
\BIBentrySTDinterwordspacing

\bibitem{st-msmarco-distilbert-base-v3}
\BIBentryALTinterwordspacing
------. (2022) msmarco‑distilbert‑base‑v3. [Online]. Available: \url{https://huggingface.co/sentence-transformers/msmarco-distilbert-base-v3}
\BIBentrySTDinterwordspacing

\bibitem{hkunlp-instructor-large}
\BIBentryALTinterwordspacing
{HKUNLP}. (2023) instructor‑large. [Online]. Available: \url{https://huggingface.co/hkunlp/instructor-large}
\BIBentrySTDinterwordspacing

\end{thebibliography}

\vspace{12pt}

\end{document}